\documentclass{article}
\usepackage{spconf,amsmath,graphicx}
\usepackage[center]{caption}
\newcommand{\norm}[1]{\left\lVert#1\right\rVert}

\title{Pretraining Strategies, Waveform Model Choice, and Acoustic Configurations for Multi-Speaker End-to-End Speech Synthesis}
\name{Erica Cooper, Xin Wang, Yi Zhao, Yusuke Yasuda, Junichi Yamagishi}
\address{National Institute of Informatics, Japan}
\begin{document}
%\ninept
%
\maketitle
\begin{abstract}
We explore pretraining strategies including choice of base corpus with the aim of choosing the best strategy for zero-shot multi-speaker end-to-end synthesis. We also examine choice of neural vocoder for waveform synthesis, as well as acoustic configurations used for mel spectrograms and final audio output.  We find that fine-tuning a multi-speaker model from found audiobook data that has passed a simple quality threshold can improve naturalness and similarity to unseen target speakers of synthetic speech.  Additionally, we find that listeners can discern between a 16kHz and 24kHz sampling rate, and that WaveRNN produces output waveforms of a comparable quality to WaveNet, with a faster inference time.
\end{abstract}

\begin{table*}[t]
\small
  \centering
  \begin{tabular}{llllllll}
  \hline
    Corpus & Speakers & Utts & Vocab size & Avg utt length & Domain & Recording quality & SNR  \\
    \hline
    VCTK & 99 & 40,373 & 5,725 & 38.3 & News \& Rainbow  & Studio & 99.8  \\
    Nancy & 1 & 11,092 & 17,894 & 82.3 & News & Studio & 92.0  \\
    LJSpeech & 1 & 12,099 & 14,696 & 99.8 & Audiobooks & High quality& 37.0  \\
    LibriTTS clean-360 & 854 & 112,038 & 67,837 & 97.6 & Audiobooks & Home (low WER) & 39.5  \\
    LibriTTS other-500 & 1148 & 199,257 & 88,253 & 89.3 & Audiobooks & Home (high WER) & 38.0 \\
    \hline
  \end{tabular}
  \caption{Information about training sets of various corpora used for model initialization and training.  Vocabulary size is number of unique tokens, and average utterance length is number of characters.}
  \label{tab:1}
\end{table*}

\section{Introduction}
\label{sec:intro}

The end-to-end approach has brought text-to-speech (TTS) synthesis into a new era in which a very high quality of synthesis can be achieved \cite{wang2017tacotron,ping2018clarinet,shen2018natural}, while also incorporating the flexibility to model different factors such as speaker identity \cite{park2019multi,jia2018transfer}, style \cite{wang2018style}, and other characteristics.  This model-based approach also allows for the inclusion and combination of diverse sources of data containing speech from a large variety of speakers in different or non-ideal recording conditions.  However, end-to-end multi-speaker TTS is very data-hungry and hence model {\em pretraining} is very important in practice.  In this paper we explore the use of different corpora for pretraining a multi-speaker synthesis model.  We also revisit some basic choices such as acoustic configurations and type of neural vocoder.  We study the effects that these choices have in the setting of Tacotron-based \cite{wang2017tacotron} zero-shot multi-speaker synthesis \cite{jia2018transfer,cooper2020zeroshot}, and in particular we study the effects these choices have on naturalness and speaker similarity.

%% TODO correlation of sentence length somehow with ali.err.rate?

Aspects of the training corpora such as number of speakers, domain of the utterances, and quality of the recording conditions can all have an effect on the final model output quality.  Text-to-speech synthesis typically requires large quantities of studio-quality recordings of professional speakers, but if we are able to relax these requirements, then our TTS models may benefit from a more diverse choice of data sources, with greater speaker variety, larger vocabulary, and different styles or domains.  In fact, the use of both high and low quality data together can provide the benefits of both.  This has been well-studied in the statistical parametric setting: in \cite{yamagishi2008robustness}, noisy target speaker data was used to adapt a Hidden Markov Model voice trained on clean multi-speaker data, resulting in a high-quality adapted voice despite the noise in the adaptation data.  Conversely, \cite{yamagishi2010thousands} trained a base model on noisy speech and adapted to clean recordings, finding that the inclusion of noisy data resulted in improvements over just using the clean data by itself, when only a small amount of clean data is available.  \cite{birusubset} extended this approach to the low-resource language of Amharic, showing that training a model on lower-quality data collected for speech recognition and adapting it to even as little as ten minutes of higher-quality audiobook data produced better synthesis than using either dataset by itself.  More recently, in the end-to-end setting, \cite{hu2019neural} used low-quality recordings to conduct fine-tuning-based speaker adaptation.  Variational autoencoder based clean speech and noise factorization has also been proposed as a way to make use of both noisy and clean data together, with \cite{8683561} using artificially-corrupted speech and \cite{hsu2019hierarchical} using real noisy speech, to create speaker-adapted TTS voices with clean output conditions.  %In this work, we are particularly interested in the {\em zero-shot} adaptation scenario \cite{jia2018transfer,cooper2020zeroshot}, in which a separately-trained speaker encoding network is used to produce speaker embeddings, which are then used as speaker information during TTS training. This approach has the advantage that only a very small amount of untranscribed target speaker data is required, and also no additional model training or fine-tuning is necessary to adapt to a new speaker.  However, this approach tends to suffer from overfitting to speakers seen during training, so we are especially interested in studying these tradeoffs of data quantity and quality with respect to speaker similarity of unseen speakers.  

In addition to the choice of data for the base model, the training strategy is also important and can make a difference to the quality of the output.  For instance, \cite{zhao2020transferring} found that for copy synthesis of music, starting with vocoder models that had originally been trained on speech data and then fine-tuning using music data worked better even than training only on music data from scratch, indicating that useful additional information for music signals can be learned from speech.  For multi-speaker text-to-speech synthesis, \cite{cooper2020zeroshot} found that warm-starting the multi-speaker model from a high-quality single-speaker pretrained model can greatly reduce training time as compared to training a multi-speaker model from scratch, and that the warm-starting approach also provides the benefits of a larger vocabulary.

 %* probably have to cite our recent IS paper

 %* can cite sp18 paper on how audiobook speech and broadcast news differs from traditional TTS corpora
 
In addition to pretraining strategies and data, we are also interested in the effect of our choice of acoustic configurations and vocoder.  Previous studies have investigated different configurations for text-to-speech synthesis in terms of acoustic parameters, type of features, training approach, and other choices.  In \cite{yamagishi2010simple}, not only is it pointed out that speech waveforms sampled at 16kHz have a less clear sound as compared to higher sampling rates, but it was also found that training models on data of a higher sampling rate results in HMM-based synthesis with better speaker similarity, indicating that the higher sampling rate enables better feature extraction.  %In \cite{Fong2019}, the differences between phone input and letter input are quantified with respect to the number of out-of-vocabulary words in the training set, and \cite{yasuda2020investigation} found that letter-based input produces comparable synthesis quality to phoneme-based input in terms of correct pronunciations, as long as the encoder has enough capacity.  -- we're not studying this
Multiple studies have also been conducted that compare different popular vocoders of the time in a listening test using copy-synthesized samples, such as \cite{airaksinen2012analysis,hu2013experimental,govalkar2019comparison}, with \cite{govalkar2019comparison} most recently finding that a WaveNet vocoder \cite{wavenet} receives the best MUSHRA scores (at the expense of slow inference).

In this paper, we explore training data and strategy, acoustic configurations, and choice of vocoder for the case of multi-speaker Tacotron speech synthesis.  We take particular interest in speaker similarity of unseen speakers in the case of zero-shot adaptation. In this approach, a separately-trained speaker encoder network is used to produce speaker embeddings, which are then input as speaker information during TTS training. This approach has the advantage that only a very small amount of untranscribed target speaker data is required, and also no additional model training or fine-tuning is necessary to adapt to a new speaker. However, previous work in this area \cite{jia2018transfer,cooper2020zeroshot} has reported that synthesis of unseen speakers tends to have lower similarity to the target speaker than synthesis of speakers who were seen during training, indicating that overfitting to training speakers may be taking place.   We study the effects of warm-starting and fine-tuning multi-speaker Tacotron models using different base corpora, investing the effects of base corpus quality, number of speakers, and vocabulary size.  We compare 16kHz and 24kHz synthesis output with the hypothesis that the higher sampling rate will not only produce better audio quality but also better represent speaker characteristics, as was the case for HMM-based synthesis in \cite{yamagishi2010simple}.  Finally, we compare WaveNet \cite{wavenet} and WaveRNN \cite{kalchbrenner2018efficient} neural vocoders to learn whether one of these is better for our task.  We conduct a large-scale listening test to subjectively evaluate the effects of these decisions.

%% our interspeech paper -- lots more speakers didn't help speaker similarity

%% TODO analyze vocabulary size. maybe try ASR ?

\section{Base Corpora and Training Strategy}
\label{sec:training}

Our aim is to train a multi-speaker Tacotron model based on the VCTK corpus \cite{vctk}, that has high synthesis quality and that also generalizes well to unseen speakers in the zero-shot adaptation case.  VCTK contains studio-quality recordings of read speech from over 100 speakers of different English dialects.  We try the strategy of training from scratch with VCTK only, and also, following \cite{cooper2020zeroshot}, warm-starting from the studio-quality, single-speaker Nancy corpus (Blizzard 2011 \cite{blizzard2011}).  We also try warm-starting from the LJSpeech corpus \cite{ljspeech17} because this corpus is in the public domain, and is comparable to the Nancy corpus in terms of number and length of utterances as well as vocabulary size.  Although the LJSpeech recordings are of relatively high quality, they nevertheless contain some audible reverberation and compression artifacts, and we wish to explore this tradeoff between lower audio quality vs.\ public availability of the data with respect to the final TTS output quality.

Since \cite{jia2018transfer} found that including more speakers during initial training from a larger corpus, such as LibriTTS \cite{libritts}, helped to mitigate the overfitting to seen speakers, we tried one more pretraining strategy: warm-start a multi-speaker LibriTTS model on an LJSpeech single-speaker model, and then fine-tune the multi-speaker model to VCTK.  LibriTTS contains disjoint training subsets divided by quality of the data as measured by word error rate of automatic speech recognition; we try our pretraining strategy with both the `clean-360' and `other-500' sets separately, in order to further explore the effects of data quality on model pretraining.

Training from scratch on VCTK data takes about four days until convergence.  The Nancy and LJSpeech base models were each trained for about 5 days, the LibriTTS warm-started training ran for 3 days, and additionally one more day of training with the VCTK data was required for the warm-started and fine-tuned models to converge.

We hold out four VCTK speakers for validation, and four speakers for a test set, balancing for gender and dialect.   Details of the training sets of the various corpora that we used can be found in Table \ref{tab:1}.  We also report estimated signal-to-noise ratio (SNR) based on the WADA SNR algorithm \cite{kim2008robust}, as an objective measure of corpus quality.

\section{Sampling rate and shared acoustic configurations}
\label{sec:sr}

With the aim to produce high-quality 24kHz output, we choose the following acoustic configurations: All training audio is resampled to 24kHz, and then 80-dimensional mel spectrograms are extracted with a 50ms frame length, 12ms frame shift, FFT size of 2048, and minimum and maximum frequency of the mel filterbank chosen as 0Hz and 8kHz, respectively.  We chose the 8kHz maximum frequency to enable better compatibility of mel spectrograms (same frequency resolution) with corpora that are lower sampling rates such as 16kHz.  Since higher frequencies depend mainly on lower ones, we expect that the waveform models will be capable of learning to predict these higher frequencies.  We use these configurations for both Tacotron and for waveform models. 

To enable a comparison with \cite{cooper2020zeroshot} and also to determine whether the sampling frequency of the output makes an audible difference to listeners, we also compare the following configurations: we train Tacotron on 80-dimensional mel spectrograms extracted from 48kHz audio (the original sampling rate of VCTK), with frame length of 50ms, frame shift of 12.5ms, FFT size of 4096, and minimum and maximum frequency of the mel filterbank chosen as 0Hz and 24kHz, respectively.  Then, we use a WaveNet that takes these mel spectrograms as input, and outputs 16kHz waveforms.  Higher frequency information is lost, but inference time is faster.

\section{Vocoder}
\label{vocoder}

We compare two popular neural vocoders, WaveNet \cite{wavenet} and WaveRNN \cite{kalchbrenner2018efficient}.  WaveNet is a dilated fully-convolutional auto-regressive neural network that generates waveforms sample by sample, based on previous samples.  Conditioned on text, it can be used by itself as a speech synthesizer, or conditioned on mel spectrograms, it can be used as a neural vocoder.  Because of its auto-regressive prediction, generation with WaveNet is notoriously slow.  WaveRNN is a single-layer recurrent neural network that was designed to match the quality of WaveNet while providing a much faster inference time.  While \cite{kalchbrenner2018efficient} reports no significant difference in quality compared to WaveNet, \cite{govalkar2019comparison} do report a significant difference in some cases, possibly due to implementation differences from the original paper to the open-source code that they used.  We are interested to learn whether there are quality differences for our particular use case of multi-speaker synthesis, using open-source implementations, that justify the slower inference time of WaveNet.

\begin{figure*}[!t]
\centering
{\includegraphics[width=1.0\textwidth]{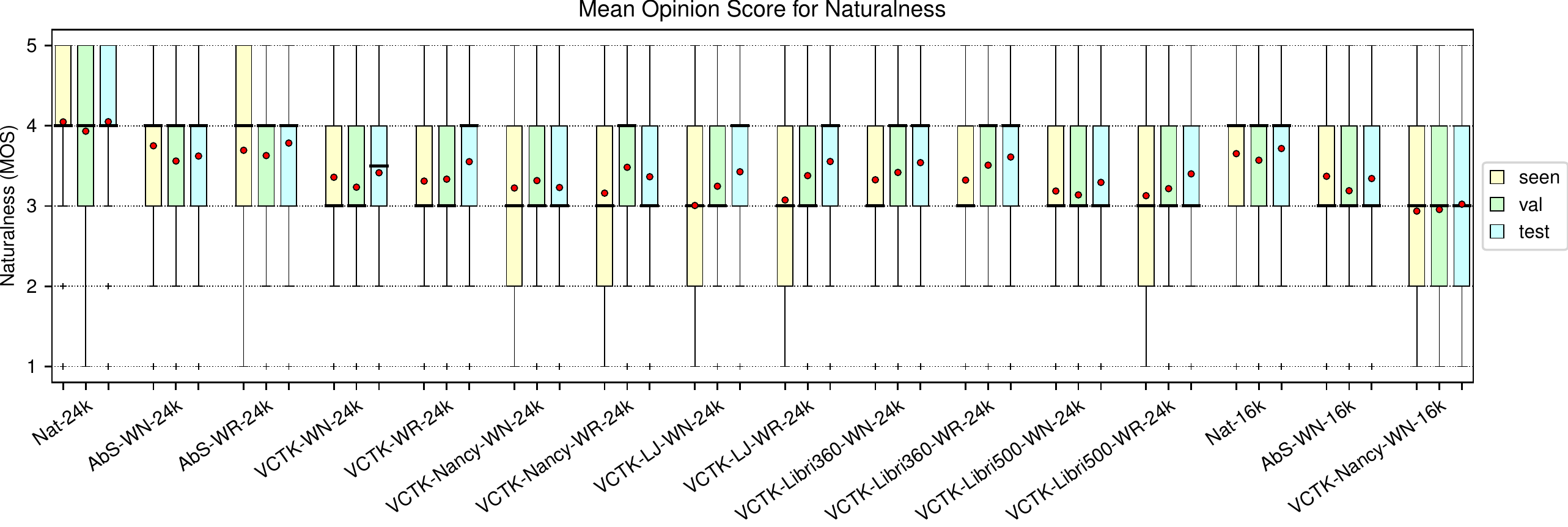}
}

{\includegraphics[width=1.0\textwidth]{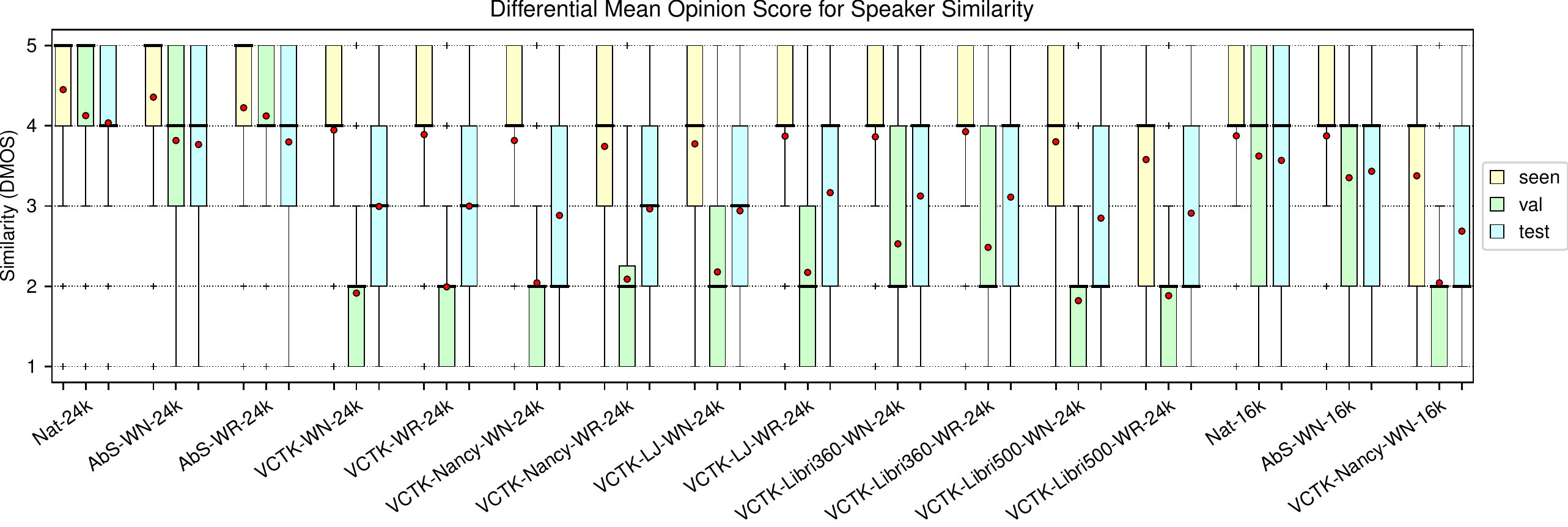}
}
\vspace{-3mm}
\caption{MOS and DMOS results for listening test} 
\label{fig:results}
\vspace{-2mm}
\end{figure*}

\section{Experiments}
\label{sec:experiments}

We use the authors' open-source implementation in \cite{cooper2020zeroshot} for our multi-speaker Tacotron-based synthesis, as well as their pre-trained learnable dictionary encoding (LDE) based speaker encoder model for speaker embeddings.  Our models use letter input and they output 80-dimensional mel spectrograms.  We convert mel spectrograms into 24kHz audio waveforms using open-source implementations of WaveNet\footnote{https://github.com/nii-yamagishilab/project-CURRENNT-scripts} and WaveRNN\footnote{https://github.com/mkotha/WaveRNN}.

For 24kHz models, we use the configurations described in Section \ref{sec:sr}.  We also include natural audio at 24kHz (\textbf{Nat-24k}) as well as copy-synthesized audio of both WaveNet and WaveRNN (Analysis by synthesis; \textbf{AbS}).  We train Tacotron models with all of the configurations described in Section \ref{sec:training}:  VCTK from scratch (\textbf{VCTK}), VCTK warm-started from Nancy (\textbf{VCTK-Nancy}), VCTK warm-started from LJSpeech (\textbf{VCTK-LJ}), and VCTK fine-tuned from LibriTTS (clean-360 and other-500, which were warm-started from LJSpeech) (\textbf{VCTK-Libri360, VCTK-Libri500}).  Following \cite{cooper2020zeroshot}, the VCTK stage of the training is gender-dependent.  For each of these models, we synthesize waveforms using both WaveNet (\textbf{WN}) and WaveRNN (\textbf{WR}), each trained on the same VCTK training set that we used for Tacotron.

We also included speech samples using the 16kHz output settings for comparison: natural speech (\textbf{Nat-16k}), copy-synthesized speech using WaveNet (\textbf{AbS-WN-16k}), and VCTK multi-speaker Tacotron warm-started from a single-speaker model trained on Nancy data, with WaveNet vocoder (\textbf{VCTK-Nancy-WN-16k}).

We conducted a crowdsourced listening test using 300 samples from each of our 16 systems, 100 samples each from four VCTK seen speakers, validation set speakers, and completely unseen test set speakers.  All sentences were text that was unseen during training.  The listening test consisted of sets of 50 samples each, distributed such that each set contains utterances from all systems.  Each listener was permitted to evaluate up to 10 sets, and we aimed for coverage of 5 ratings per utterance.  215 unique listeners participated, each completing 2 sets on average.  Participants were asked for a mean opinion score (MOS) of how natural the audio sounds on a scale of 1 to 5, and a differential MOS (DMOS) rating how similar the utterance sounds to a reference sample from the target speaker, on a scale from 1 (definitely a different speaker) to 5 (definitely the same speaker).  Results from our listening test can be found in Figure \ref{fig:results}.

\section{Results}
\label{sec:results}

We measure statistical significance using a Mann-Whitney U test, following \cite{rosenberg2017bias}, at a level of p\textless  0.01.

\subsection{Acoustic Configurations}

The 24kHz audio for natural speech was significantly preferred over 16kHz in all cases, including for speaker similarity.  The same significant preference was observed for the WaveNet copy-synthesized speech.  In the case of text-to-speech synthesis (VCTK-Nancy-WN-16k vs.\ VCTK-Nancy-WN-24k), naturalness is significantly improved by the 24kHz rate in all cases, and speaker similarity is significantly improved for seen speakers.  The improvements over the 16kHz output demonstrates that the waveform models successfully learned to predict the higher-frequency information.

\subsection{WaveNet vs.\ WaveRNN}

For analysis by synthesis of seen speakers, WaveNet was slightly but not significantly preferred over WaveRNN for both naturalness and similarity.  For unseen validation and test set speakers, WaveRNN was preferred (significantly for naturalness of test set and speaker similarity of validation set; not significantly otherwise).  In the text-to-speech case, there were no significant differences except for naturalness on the validation set of the VCTK model fine-tuned from Nancy, favoring WaveRNN.  So, it can be concluded that WaveRNN's faster inference time does not harm output quality, and may even improve it slightly.

\vspace{-3mm}
\subsection{Best pretraining strategy}

\noindent\textbf{Warm-starting from Nancy vs.\ LJSpeech:} For speaker similarity, there is no significant difference between the models warm-started from Nancy and from LJSpeech.  For naturalness, there was a significant preference for the Nancy-based model for seen speakers (with WaveNet vocoder), and for the LJSpeech-based model for test set speakers (both vocoders).  Considering that we are mainly interested in improving synthesis for {\em unseen} speakers, we can conclude that warm-starting using the freely-available public LJSpeech corpus is comparable to or better than using the more restrictively-licensed Nancy data.

%TODO why the opposite results for seen vs. test?  (since it's only one vocoder, I have doubts that the Nancy preference is "really" significant)  the sentences in each set are different, so maybe test set is somehow better matched to the sentences in LJ, and the val set sentences are better matched to Nancy.

\noindent\textbf{Fine-tuning from LibriTTS clean-360 vs.\ other-500:} 
We consider the tradeoffs between using noisier data with a larger number of speakers and vocabulary size vs.\ data with slightly fewer speakers and vocabulary but higher audio quality.  We can see that fine-tuning from the cleaner data is always better, significantly so in most cases (with exceptions for both naturalness and similarity of seen speakers using WaveNet, and similarity of test speakers using WaveRNN).  

\noindent\textbf{Best and worst models:} The best model for seen speakers was the one that was trained from scratch.  This is not surprising since this model spends the most time learning from training-set VCTK speakers and thus overfits to them.  For naturalness of unseen speakers, the best model is VCTK-Libri360-WR-24k.  For similarity of unseen speakers, the best models were VCTK-Libri360-WN-24k (for validation set speakers; the WaveRNN variant of the same model was not significantly worse) and VCTK-LJ-WR-24k (for test set speakers; the models fine-tuned from LibriTTS clean-360 were also not significantly worse.)  Consistently, the worst models for speaker similarity were the ones fine-tuned from LibriTTS other-500, indicating that despite being the corpus with the largest number of speakers, the quality of the data is too low to improve the modeling of the speaker space.
 \vspace{-3mm}
\subsection{Alignment errors}

Alignment error rate is the percent of synthesized utterances that contain any kind of error related to the alignment predicted by the attention mechanism, such as incomplete synthesis, overestimated synthesis, or discontinuities.  We computed alignment error rate for all synthesis samples of all 24kHz Tacotron models; results are reported in Table \ref{tab:3}.

\begin{table}[t]
\small
  \centering
  \begin{tabular}{lllll}
  \hline
    Model & Seen & Val & Test & Avg  \\
    \hline
    VCTK          & 2.0\% & 0.5\% & 1.0\% & 1.2\% \\
    VCTK-Nancy    & 7.0\% & 2.5\% & 2.5\% & 4.0\% \\
    VCTK-LJ       & 8.5\% & 4.0\% & 4.0\% & 5.5\% \\
    VCTK-Libri360 & 2.5\% & 3.5\% & 4.0\% & 3.3\% \\
    VCTK-Libri500 & 6.5\% & 5.0\% & 2.0\% & 4.5\% \\
    \hline
  \end{tabular}
  \caption{Alignment error rates computed on synthesized VCTK sample utterances}
  \label{tab:3}
\end{table}

It can be observed that the model trained from scratch on VCTK has the fewest alignment errors, demonstrating that training on target domain data for more time results in better learning of alignments in that domain.  Of the other training strategies, fine-tuning from LibriTTS clean-360 has the lowest alignment error rates, showing the benefits of training on larger amounts of more diverse and relatively clean data.

Since VCTK utterances are relatively short, we were also interested to learn which models were most capable of synthesizing longer utterances, so we selected the 10 longest utterances from the Nancy test set and synthesized them with each model in the ``voice'' of one seen speaker and one unseen (test set) speaker.  Results are reported in Table \ref{tab:longsentences}.

\begin{table}[t]
\small
  \centering
  \begin{tabular}{llll}
  \hline
    Model & Seen & Unseen & Avg  \\
    \hline
    VCTK          & 60\% & 80\% & 70\% \\
    VCTK-Nancy    & 90\% & 60\% & 75\% \\
    VCTK-LJ       & 80\% & 90\% & 85\% \\
    VCTK-Libri360 & 100\% & 20\% & 60\% \\
    VCTK-Libri500 & 80\% & 30\% & 55\% \\
    \hline
  \end{tabular}
  \caption{Alignment error rates computed for synthesis of very long sample utterances}
  \label{tab:longsentences}
\end{table}

It is apparent that these very long utterances are challenging for these models to synthesize, as all of the alignment error rates are very high.  This can possibly be mitigated in future work by incorporating recent techniques for improving alignment of longer utterances, such as \cite{battenberg2020location}.  Interestingly, the model fine-tuned from LibriTTS other-500 has the fewest errors on average, indicating that this training approach does allow the model to learn something more about synthesis of very long sentences, despite producing the worst-rated audio.
\vspace{-4mm}
\subsection{Objective measures of speaker similarity}

\begin{figure}[tb] 
    \centering
    \includegraphics[width=1.0\linewidth]{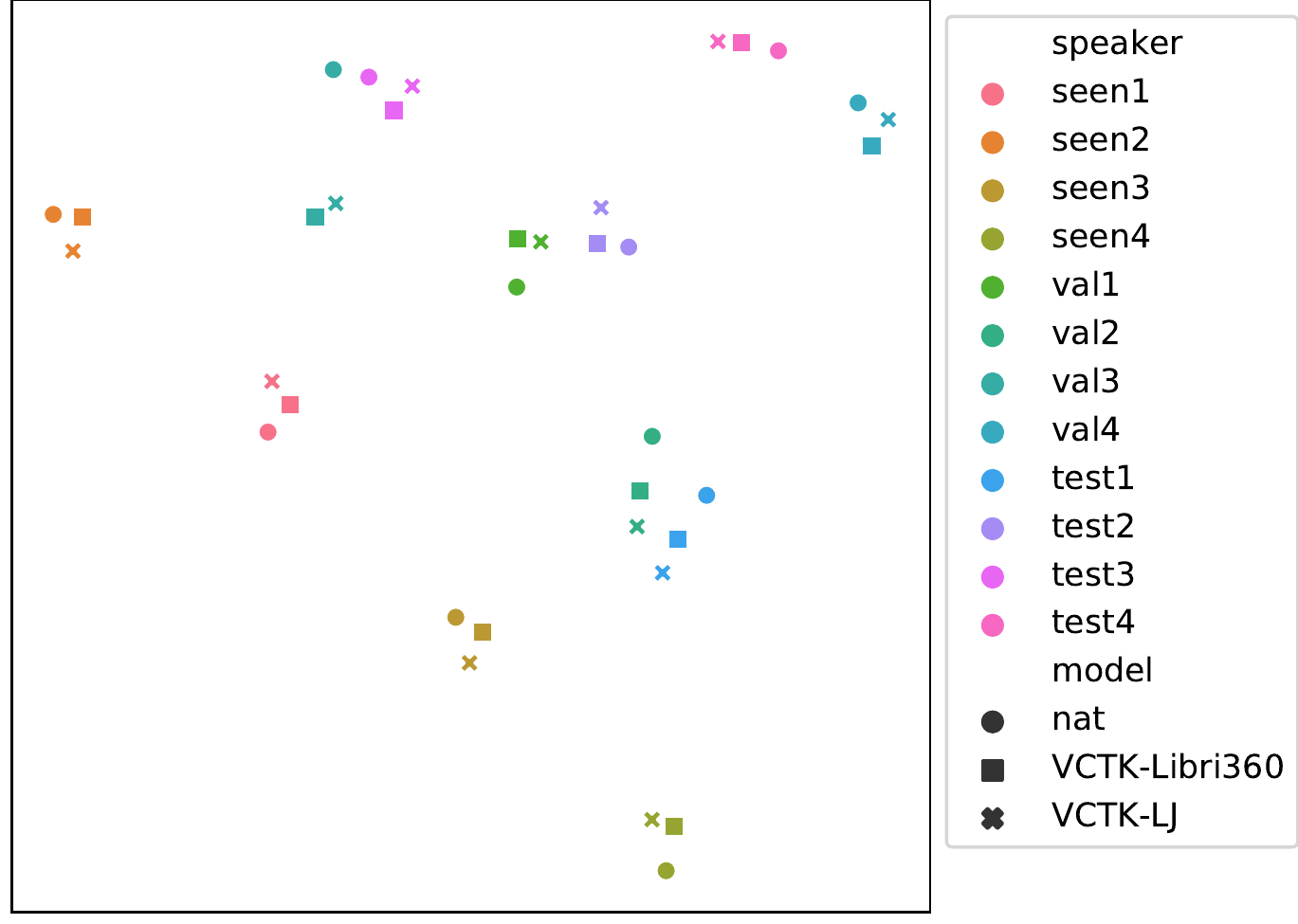}
    \vspace{-3mm}
    \caption{t-SNE plot of x-vectors from natural speech, VCTK-LJ, and VCTK-Libri360}
    \label{fig:tsne}
    \vspace{-5mm}
    \end{figure}

We can observe how synthesized speech becomes more similar to the target speakers' speech in speaker space by measuring cosine similarities of embeddings extracted from synthesized speech, compared to embeddings extracted from original speech.  Despite the fact that we used LDE-based embeddings for training our TTS models, we use x-vectors \cite{snyder2018x} for the objective evaluation in order to measure in a way that is independent of how we defined speaker characteristics during Tacotron training.  Cosine similarity is a standard measure of similarity of embedding vectors that is commonly used in the field of automatic speaker verification; it is defined as $cos\_sim(A,B) = A \cdot B/\norm{A}\norm{B}$ and it ranges from -1 to 1, with 1 being most similar.  For simplicity, we report results from synthesis using WaveRNN only, in Table \ref{tab:cossim}.  We can observe a strong Pearson product-moment correlation between the cosine similarities and the subjective measures of speaker similarity from our listening test, with r=0.89.  To further illustrate the speaker space, we also generated a t-SNE plot (Figure \ref{fig:tsne}) of x-vectors for each speaker's natural speech and synthesized speech from models VCTK-LJ and VCTK-Libri360, in which we can observe that the VCTK-Libri360 embeddings generally move closer towards the true speech.

\begin{table}[t]
\small
  \centering
  \begin{tabular}{lllll}
  \hline
    Model & Seen & Val & Test & Avg  \\
    \hline
    AbS           & 0.98 & 0.98 & 0.98 & 0.98 \\
    VCTK          & 0.82 & 0.50 & 0.53 & 0.62 \\
    VCTK-Nancy    & 0.77 & 0.48 & 0.53 & 0.59 \\
    VCTK-LJ       & 0.78 & 0.53 & 0.52 & 0.61 \\
    VCTK-Libri360 & 0.82 & 0.58 & 0.57 & 0.65 \\
    VCTK-Libri500 & 0.79 & 0.42 & 0.52 & 0.58 \\
    \hline
  \end{tabular}
  \caption{Cosine similarities to seen and unseen speakers for different model training strategies}
  \label{tab:cossim}
\end{table}

\vspace{-3mm}
\subsection{Automatic measures of intelligibility}

In order to assess the intelligibility of the synthesized speech, we submitted a sample of the test utterances to the IBM Watson speech-to-text API\footnote{https://www.ibm.com/cloud/watson-speech-to-text}, which has been shown to correlate well with human judgments of intelligibility of synthesized speech obtained in a transcription task \cite{cooper2017utterance}.  The API returns a hypothesized transcription, and we measured word error rate (WER) between the hypothesized transcription and the ground-truth text from which the utterance was originally synthesized.  WER is computed as the Levenshtein distance between the two word sequences divided by the length of the ground-truth sequence.  Averaged WER results for each system are reported in Table \ref{tab:5}.  The 6\% WER of natural speech represents a top line of how well the system can recognize real speech.  We can observe some degradation simply from the AbS.  Interestingly the WERs do not correlate with alignment error rates, indicating that there may be pronunciation issues which are distinct from the alignment errors.

\vspace{-2mm}
\section{Conclusions}
\label{sec:conclusions}

We have observed that the LibriTTS clean-360 dataset is effective for pretraining a multi-speaker Tacotron model for VCTK.  We particularly note a large difference in quality between models fine-tuned using LibriTTS clean-360 and other-500, indicating that the lower recording quality of the latter set outweighs any benefits gained from more speakers and a larger vocabulary.  We also observe that the freely-available LJSpeech corpus is a suitable substitute for the more restrictively-licensed Nancy corpus for the purpose of warm-starting the multi-speaker model.  Furthermore, WaveRNN can produce high-quality waveforms for our task with a faster inference time than WaveNet, and that a 24kHz sample rate is worthwhile to generate since listeners prefer it.  We identify alignment errors especially for longer synthesized utterances as well as potential intelligibility issues, even for our best models, which is an area for improvement in future work.  Overall, our results indicate that pretraining of end-to-end multi-speaker synthesis models may be done using home-recorded audio data instead of studio recorded high-quality audio data, and this is good news for speech synthesis in under-resourced languages or dialects.

\begin{table}[t]
\small
  \centering
  \begin{tabular}{lllll}
  \hline
    Model & Seen & Val & Test & Avg  \\
    \hline
    Nat           & 6\% & 9\%   & 3\%  & 6\% \\
    AbS           & 8\% & 11\%  & 11\% & 10\% \\
    VCTK          & 13\% & 13\% & 17\% & 14\% \\
    VCTK-Nancy    & 19\% & 18\% & 20\% & 19\% \\
    VCTK-LJ       & 12\% & 10\% & 14\% & 12\% \\
    VCTK-Libri360 & 16\% & 10\% & 18\% & 15\% \\
    VCTK-Libri500 & 16\% & 9\%  & 14\% & 13\% \\
    \hline
  \end{tabular}
  \caption{Word error rates from automatic speech recognition}
  \label{tab:5}
\end{table}

\vspace{-2mm}
\section{Acknowledgments}
\label{sec:acknowledgments}

This work was partially supported by a JST CREST Grant (JPMJCR18A6, VoicePersonae project), Japan,  MEXT KAKENHI Grants (16H06302, 18H04120, 18H04112, 18KT0051, 19K24372, 19K24373, 19K24371), Japan.  The numerical calculations were carried out on the TSUBAME 3.0 supercomputer at the Tokyo Institute of Technology.

\bibliographystyle{IEEEbib}
\bibliography{strings,refs}

\end{document}